# Exact clustering in linear time

Jonathan A. Marshall[1]*, Lawrence C. Rafsky[1]
2017-02-17. Submitted for publication.

[1]Acquire Media, 3 Becker Farm Road, Roseland, NJ 07068, USA.

*Correspondence to: jmarshall@acquiremedia.com.

**Abstract**: The time complexity of data clustering has been viewed as fundamentally quadratic, slowing with the number of data items, as each item is compared for similarity to preceding items. Clustering of large data sets has been infeasible without resorting to probabilistic methods or to capping the number of clusters. Here we introduce MIMOSA, a novel class of algorithms which achieve linear-time computational complexity on clustering tasks. MIMOSA algorithms mark and match partial-signature keys in a hash table to obtain exact, error-free cluster retrieval. Benchmark measurements, on clustering a data set of 10,000,000 news articles by news topic, found that a MIMOSA implementation finished more than four orders of magnitude faster than a standard centroid implementation.

**Summary:** Big-data computations that would have taken years can now be done in minutes.

## INTRODUCTION

Data clustering techniques are widely used in computational data science. With increasing data capacities and speeds in computing, scientists in many domains (*1, 2, 3, 4, 5, 6*) seek to perform clustering on ever-larger "big data" sets.

Clustering may entail comparing data items to one another along several dimensions, and assigning similar data items to the same group. With large data sets, similarity computation becomes slow and expensive, as each data item is compared to a large number of other data items. The time complexity of similarity clustering has been viewed as fundamentally $O(n^2)$: quadratic in the number of data items.

Even with aggressive techniques such as probabilistic algorithms (*3, 4, 5, 6, 7, 8, 9, 10*), partitioning (*3, 4, 10, 11, 12*), and parallelization, comparing similarity between the items in a large data set can require a prohibitive amount of computation (*2, 3, 5, 7, 8, 9*), or can yield generally inferior clustering (*13, 14*). The *k*-means algorithm limits the comparisons of each item to *k* cluster centroids, resulting in O($nk$) time complexity, for a small, fixed value of *k* determined in advance. Large data sets may have many clusters; limiting the number to a fixed *k* may result in inadequate cluster quality for certain applications.

Similarity clustering in linear or near-linear time can be obtained via probabilistic clustering algorithms, such as MinHash (*4, 7, 8, 9*) methods – but at the cost of admitting errors in retrieval, such as false negatives, in which the algorithm may (with small probability) erroneously omit certain cluster members during cluster retrieval. An omission may be tolerable in some application domains (e.g., document deduplication, advertisement targeting), but unacceptable in others (e.g., medical diagnosis, scientific analysis, engineering designs) – which may require or prefer an error-free, or exact, clustering method rather than a probabilistic, or approximate, one.

Unlike earlier methods, MIMOSA (Mark-In, Match-Out Similarity Analysis) algorithms perform exact similarity clustering in $O(n)$ time (linear time complexity in the number of data items) – not suffering probabilistic errors, nor capping the number of clusters. Because a MIMOSA algorithm takes about the same time to process the millionth data item as it takes to process the first, it performs clustering faster than other exact methods when the number of data items is large.

## Data item signatures

MIMOSA algorithms are signature-based. Each data item has a signature $S_i = \{S_{i1}, ..., S_{in_i}\}$: a limited-size set of elements that describe the data item, so that the signatures of similar data items may have one or more elements in common. MIMOSA finds data items whose signatures are similar, and clusters them accordingly.

For example, in a news analysis application where each data item is a news article, a signature might be a set of keywords denoting the most important people, companies, and events in the article. An article of 700 words, entitled "School, infrastructure bond measures fill U.S. ballots," might have signature BALLOT-BOND-BORROW-CALIFORNIA-INFRASTRUCTURE-MEASURE-MUNICIPAL-SCHOOL-TAX-TRANSIT-VOTE-YIELD. Each element is chosen or derived for high informational value. Terms of lower value, such as common stopwords ("the") or words appearing infrequently in the article ("airport") are typically omitted from a news article signature. Articles whose signatures share several elements – i.e., cover the same news topic – can belong to the same cluster.

In other example applications, a signature can describe the expressed proteins from a gene, chemical activity measurements, a user's web browsing behavior, a mailing address for marketing, patient medical symptoms, a business's credit history, psychological, demographic, or census survey entries, or sensor readings from a scientific apparatus or industrial machine.

## Structure of MIMOSA algorithms

A MIMOSA run is preconfigured by specifying a similarity measure $s()$, a minimum similarity threshold value $\theta$, and a list $A$ of the size values that are allowed or expected for signatures. When $s(X,Y)$ meets or exceeds $\theta$, then $X$ and $Y$ are said to be similar to each other. One popular and useful similarity measure is Jaccard similarity, $s(X,Y) \equiv |X \cap Y| / |X \cup Y|$, in which the pairwise similarity score depends on the sizes of both the intersection (overlap) and the union of the two signatures $X$ and $Y$.

A partial signature is a subset of the elements of a signature. For the news article example above, one of the partial signatures is BOND-CALIFORNIA-SCHOOL-TAX-VOTE. This example partial signature has size 5, the number of its elements. A signature of size $n$ has at most $2^n - 1$ partial signatures.

From each signature $S_i$, a MIMOSA algorithm derives a set of partial signatures of certain sizes:

$$C_i = \bigcup_{j \in L_{|S_i|}} \binom{S_i}{j}, \text{ where } |S_i| \text{ is the size of (number of elements in) signature } S_i.$$

The term $\binom{S_i}{j}$ refers to the set of all partial signatures of size $j$, drawn from the elements of signature $S_i$. The set $C_i$ contains the partial signatures of several sizes $j$. The size values $j$ come from a set $L_{|S_i|}$ which depends on the size $|S_i|$ of the signature $S_i$. Each such set $L_x$ is called the MinOverlap set for size $x$. The MinOverlap set $L_x$ contains the sizes of the smallest overlaps between two signatures (one of size $x$, and the other of size $y$), subject to the constraint that the two signatures are similar to each other:

$$L_x = \left\{ \left( \min_{\{Y: |Y|=y,\ s(X,Y) \geq \theta\}} |X \cap Y| \right) : |X| = x,\ y \in A \right\}.$$

For each signature size $y \in A$, there is at most one MinOverlap value in $L_x$. The set $L_x$ might not contain all size values between 1 and $|S_i|$. Hence, the number of partial signatures in $C_i$ for a given signature $S_i$ may be less than $2^{|S_i|} - 1$, but cannot exceed $2^{|S_i|} - 1$.

A MIMOSA algorithm constructs memory storage keys from the partial signatures $C_i$, so that two signatures are similar if and only if they have at least one key in common. Each key contains the elements of a partial signature, concatenated together in a sorted order, using a separator character.

MIMOSA algorithms use the keys to store and check marker values in memory. Because the number of keys per data item is upper-bounded, a MIMOSA algorithm's runtime per data item does not exceed a fixed maximum – regardless of the number of data items. This property is what makes the time complexity of MIMOSA algorithms linear in the number of data items.

During a "Match-Out" stage, MIMOSA algorithms check whether one or more keys are marked in a hash table (or any key-value data structure that allows insertion and retrieval in constant time). The number of keys checked per data item does not exceed a fixed maximum.

During a "Mark-In" stage, MIMOSA algorithms mark a set of keys from each data item in the hash table. Depending on implementation, the value of a marker may represent a simple flag to indicate just that the key is marked; or it may represent (or point to) other information, such as a signature, a cluster identifier, or a key.

The keys derived from a signature conceptually represent points in a multi-dimensional neighborhood surrounding the signature. Checking the keys for a data item conceptually represents looking outward from the data item signature to determine whether a surrounding multi-dimensional neighborhood overlaps with the marked neighborhoods surrounding other data item signatures.

MIMOSA algorithms assign the data item to a cluster based on which keys are marked or unmarked. For example, if none of the keys for a data item is marked, then the data item is not similar to any other data items. A MIMOSA algorithm may then assign the data item to a new cluster by marking its keys with a new cluster identifier. Or if a key is marked, the data item is similar to another data item. A MIMOSA algorithm may then assign it to an existing cluster by marking its keys with an existing cluster identifier.

A MIMOSA algorithm may also handle the case where keys corresponding to two or more clusters are marked. It may have conditions and rules that assign the data item to a chosen cluster, by marking its keys accordingly. It may optionally have conditions and rules for certain side effects, such as merging or splitting clusters.

The way that MIMOSA algorithms find a cluster can be described conceptually as a two-step process: traversing "outward" from the signature to its keys, and then "inward" from a matching key to a marker that represents the cluster.

## Matching and marking keys to identify similar signatures

**Figure 1** depicts a simplified example sequence of data item signatures processed by a MIMOSA algorithm. The example uses Jaccard similarity, with $\theta = 0.3$. The signatures all have 4 elements: $A = \{4\}$. Two signatures are similar if they have at least 2 elements in common.

After each data item signature is received, the MIMOSA algorithm generates keys comprising all 2-element combinations of the signature. The algorithm checks whether any of the keys is marked in the hash table. If it finds none, it generates a new cluster ID value. Or if it finds at least one of those keys marked, it obtains an existing cluster ID from the corresponding marker value.

Using the new or existing cluster ID, the MIMOSA algorithm marks the keys, writing marker values containing that cluster ID into the hash table.

Even though more and more marker values are stored into the hash table, the amount of time that the algorithm takes to check each data item and assign it to a cluster does not grow. The reason for this important characteristic is that the number of keys per data item depends on the data item itself, but not on the number of data items. The algorithm's linear time complexity allows it to run without slowing when the number of data items is large.

## Keeping track of signature size

**Figure 2** shows that the simple keys comprising a partial signature may not be enough, in certain cases. The goal is to create keys such that two signatures are similar if and only if they have at least one key in common. The figure compares simple keys (**A**) of the type used in Figure 1 with keys that contain an additional component (**B**): a numeric size value. Both parts of the figure illustrate the state of memory after three input signatures are received.

In the example, Jaccard similarity is used, $\theta = 0.4$, and $A = \{3,4\}$ (the signatures contain either 3 or 4 elements). From these conditions, it may be deduced that:

- two signatures, each comprising 3 elements, are similar to each other if they have at least 2 elements in common;

- a signature comprising 3 elements is similar to a signature comprising 4 elements if they have at least 2 elements in common; and

- two signatures, each comprising 4 elements, are similar to each other if they have at least 3 elements in common.

Consequently, the MIMOSA algorithm of Figure 2A uses partial signatures of sizes 2 and 3 for input signature A-B-C-D. The figure shows that the algorithm generated keys comprising all such partial signatures (A-B, A-C, A-D, B-C, B-D, C-D, and A-B-C, A-B-D, A-C-D, B-C-D) and marked them in the hash table at $t = 1$. Likewise, for input signature E-F-G, it uses partial signatures of size 2. It generated keys comprising all such partial signatures (E-F, E-G, and F-G) and marked them at $t = 2$.

Partial signatures of size 1 are not used because, under the given conditions, two signatures of sizes 3 or 4 cannot be similar if they share only 1 element in common. Partial signatures of size 4 are also not used because, under the given conditions, they are unneeded: if two similar signatures have 4 elements in common, they also have a partial signature of 3 elements in common that suffices to show that they are similar. In general, a MIMOSA algorithm may use the smallest partial signatures in common that show two signatures to be similar.

When signature A-B-E-F is received at $t = 3$, the algorithm of Figure 2A generates partial signatures of sizes 2 and 3. As shown, the keys of two of those partial signatures, A-B and E-F, match keys marked by previous inputs. One of the previous input signatures, E-F-G, is indeed similar to A-B-E-F under the given similarity measure and threshold: with $X = \{E, F, G\}$ and $Y = \{A, B, E, F\}$, $s(X, Y) = |X \cap Y| / |X \cup Y| = 2/5$, which is $\geq \theta$.

However, the simple approach leads to a problem. The other matched signature, A-B-C-D, is actually *not* similar to A-B-E-F under the given similarity measure and threshold. With $X = \{A, B, C, D\}$ and $Y = \{A, B, E, F\}$, $s(X, Y) = |X \cap Y| / |X \cup Y| = 2/6$, which is $< \theta$. In other words, the matched key A-B would falsely indicate that data item signatures A-B-C-D and A-B-E-F are similar.

The MIMOSA algorithm of Figure 2B avoids this problem by using a slightly different method of creating keys from partial signatures. This method inserts an additional value into each key: a number representing the size of a signature. During the Mark-In stage, when keys are marked in the hash table, the method uses the size of the input signature. During the Match-Out stage, when the presence of keys is checked in the hash table, the method uses the sizes of signatures that can be similar to the input signature.

Figure 2B depicts this dual-stage approach. The keys for A-B-C-D were prefixed with 4, representing the size of A-B-C-D, during the Mark-In stage when those keys were marked in the hash table. The keys for E-F-G were prefixed with 3, representing the size of E-F-G, during the Mark-In stage when those keys were marked in the hash table. These marked keys, containing the size of the input signature, are called MI (Mark-In) keys.

During the Match-Out stage for input signature A-B-E-F, the method of Figure 2B prefixes some of the keys with 3, and others with 4, depending on the size of the signatures that they are intended to match. These keys, containing the size of a signature to be matched, are called MO (Match-Out) keys. The MO keys are used for checking the hash table, but are not marked in the hash table.

During the Mark-In stage for A-B-E-F, (not shown), the method of Figure 2B will mark MI keys (all prefixed with 4) for A-B-E-F in the hash table.

Because of the differences between the two types of keys in the method of Figure 2B, the marked MI key containing A-B does *not* match the checked MO key containing A-B; the keys contain different prefixes (3-A-B versus 4-A-B). Consequently, the MIMOSA algorithm of Figure 2B avoids reporting (falsely) that signature A-B-E-F is similar to A-B-C-D.

Thus, the determination of whether two signatures are similar may depend not only on their matched partial signatures, but also on the sizes of both signatures. Figure 2B shows one way that a MIMOSA algorithm may correctly keep track of the size of a signature, and may use that size information when it checks for matching tokens from other signatures, to determine similarity without false-positive errors.

Given $s(\ )$, $\theta$, and $A$, all MI and MO size values may be precalculated once, on initialization, and stored in a table. Then the size values needed for the keys may be obtained directly from the table during MIMOSA operation on each data item.

## MATERIALS AND METHODS

Performance of a MIMOSA implementation and a control implementation of a standard centroid clustering algorithm were compared in a benchmark test. Complete software source code and data sets are provided in the Supplementary Materials. In the centroid implementation, the similarity of each signature to one designated signature (termed the centroid) of each existing cluster is computed. As additional data items are received, the number of clusters tends to increase. Consequently, as each additional signature is received, the standard centroid clustering algorithm takes longer to compare the signature to all existing cluster centroids.

Both implementations were written in the same programming language, used a single running thread, used the same Jaccard similarity measure, were run on the same computer, and received the same input data.

The input data items in this run were a set of 10,000,000 news articles from several thousand publishers, commercially distributed in January 2017. Prior to and separate from clustering, a signature was computed for each article, comprising up to 10 key terms chosen to characterize the content and topic of the article. The elements within each signature were provided in a sorted order.

In the MIMOSA implementation, the hash table was initially empty. The program formed, checked, and inserted keys in the hash table on the fly as each input signature was received. The population of keys into the hash table created a quick-lookup data structure, allowing each successive signature to be compared for similarity with all other signatures in the data structure, within a fixed, constant time.

**Performance outcome**

**Figure 3A** shows a plot of the cumulative average time (in seconds) to cluster each signature, as a function of the number of received signatures, on a log-log scale. The clustering time per signature for the MIMOSA run was constant, about 0.00030 seconds per item, regardless of the number of received items. The clustering time per signature for the standard centroid clustering algorithm grew linearly, and eventually exceeded 0.41 seconds per signature on average, reaching 0.65 seconds per signature for the last items in the run.

The centroid algorithm implementation clustered 250,000 signatures in 103,589 seconds (1 day 4 hours 46 minutes 29 seconds); at that point the test of the centroid algorithm was stopped because the results were sufficient for the benchmark. In comparison, the MIMOSA implementation clustered the 250,000 signatures in 88 seconds (1 minute 28 seconds) – that is, 1,177 times faster.

The MIMOSA implementation continued to cluster the full set of 10,000,000 signatures in 2,982 seconds (49 minutes 42 seconds).

If the growth of clustering time of the centroid algorithm is modeled as quadratic, the implementation would take an extrapolated $\left(10{,}000{,}000 \times \sqrt{103{,}589} \, / \, 250{,}000 \right)^2$ seconds, or about 5.2 years – likely too long to wait – for the centroid algorithm to cluster 10,000,000 signatures, averaging about 16.5 seconds per signature. The MIMOSA implementation would thus be about 55,000 times faster (more than 4 orders of magnitude) than the centroid implementation in clustering 10,000,000 signatures.

MIMOSA memory usage is bounded linearly in the number of data items. The total amount of memory that MIMOSA used in clustering 10,000,000 signatures grew to 61.3 gigabytes, well within the memory capacity of a single, off-the-shelf computer server.

**Figure 3B** shows a linear-scale plot of the total clustering run time for both implementations, as a function of the number of signatures received. An inset in the figure enlarges the plot on the first 100 signatures. The centroid algorithm implementation was faster than the MIMOSA implementation until about 70 signatures were clustered. The inset also reveals the linear time-complexity of MIMOSA, in comparison to the quadratic time-complexity of the standard centroid clustering method.

The timings reported in Figure 3 represent the full activity of the two implementations. For MIMOSA, this included identifying the signatures, generating partial signatures, forming keys, checking and storing keys in the hash table, and outputting the assigned cluster identifiers.

**Clustering outcome**

**Figure 4** shows histograms of the sizes of the clusters produced by the centroid implementation and the MIMOSA implementation of Figure 3. The first two panels show the distribution of cluster sizes that result from running the two implementations on 250,000 signatures. The clusters formed are identical in the two implementations – confirming that both solved exactly the same clustering task. The third panel shows the sizes of the clusters that result from continuing the MIMOSA run to 10,000,000 signatures. The clusters are generally larger, because of the greater number of data items. The shape of the distribution is similar to that of the shorter clustering run.

Each cluster created in the run represents a set of news articles that are similar to one another, in that their signatures have several elements in common. In other words, each cluster represents a news topic. A few of the clusters are large (topics with many articles); many are small. Headlines indicating the topics of large news clusters derived from these articles around January 2017 include:

- "Harsh rhetoric from Moscow over Obama hacking sanctions"
- "Israel-Palestine conflict could 'explode' under Donald Trump, Israel supporter warns"
- "Of course Mexico will not pay"

**MIMOSA variations**

Clustering methods are commonly tailored to the needs of the application at hand (*1, 2, 3, 10*). MIMOSA is a framework that supports a variety of algorithmic features. It is straightforwardly parallelizable and scalable (for example, via a map-reduce framework (*15*) in which the hash table is partitioned and distributed across multiple computers). It works with online streaming data sequences, as well as with batch data. It can be run repeatedly with different $\theta$ values to form a hierarchy of clusterings. It can check for similarity to signatures or to centroids. It supports a variety of decision rules to determine cluster membership. It can be combined with direct similarity calculations to support fractional, fuzzy, weighted, or probabilistic thresholds and elements – not just integer sizes. It can allow assignment of each data item to one cluster or to multiple clusters.

The dimensionality of MIMOSA clustering is arbitrarily high. As illustrated, the symbols representing each signature element can be any string – and there is no limit on how many such symbols may exist across all signatures. Moreover, the number of possible symbols does not need to be known at any stage. Each data item is represented by an identified signature of at most $\max(A)$ elements, which typically would correspond to the largest or most significant $\max(A)$ dimensions for each data item, as effected by the signature generation process.

MIMOSA algorithms achieve linear time-complexity and linear space-complexity by capping the number of dimensions *per data item* to $\max(A)$, while allowing any number of dimensions *per data set*. MIMOSA is suitable for clustering applications where it is allowable to limit or truncate the signature sizes at $\max(A)$, and where the values of $\max(A)$ and $\theta$ can be chosen to yield a manageable number of keys, with markers fitting in available memory.

**Concluding discussion**

It is unusual for a computational task thought to be quadratic to be reimagined as linear – especially a task as well studied and widely applied as exact clustering. As illustrated in Figure 3B, the impact of switching from quadratic to linear is not just enormous in magnitude but is also profound structurally – eliminating the compounded slowdown per data item.

In hindsight, we see that the reason standard clustering algorithms are fundamentally slow is that they spend much time checking each data item against previous items that have no chance of being similar. By limiting checks to fixed-size neighborhoods in which similarity exists, MIMOSA algorithms avoid unnecessary work.

The MIMOSA framework is expected to become helpful in areas where clustering or grouping analysis of large datasets has been hindered by slow or probabilistic methods, or by limiting the number of clusters.

## ACKNOWLEDGEMENTS

The authors are grateful to Acquire Media for support of this research. Both authors are employed by Acquire Media. Dr. Marshall is the senior author. Dr. Rafsky is a part owner of Acquire Media. The authors have submitted patent applications, pending and assigned to Acquire Media, that include material described in this paper. The software and data reported in this paper are archived at http://mimosa.acquiremedia.com/ .

## FIGURES

**Figure 1. MIMOSA clustering of 10 data item signatures.** Each panel represents the state of memory after an input signature is received. In this sketch, each signature has 4 elements; and two signatures must have at least 2 elements in common to be similar. MIMOSA marks 2-element keys. The hash table in memory starts empty at $t = 0$. At $t = 1$, signature A-B-C-D is received. MIMOSA finds none of its 2-element keys marked in the hash table, so it assigns cluster ID #1, by marking them in the hash table with markers indicating #1. At $t = 2$, MIMOSA finds no matching keys for D-E-F-G, so assigns it to cluster #2. At $t = 3$, MIMOSA finds one matching key (E-G), so assigns A-E-G-H to cluster #2. At $t = 4, 5, 6, 7, 8, 9, 10$, MIMOSA receives signatures B-C-E-I, C-F-H-J, D-E-J-K, C-G-K-L, D-H-I-L, C-I-M-N, C-F-H-O, and assigns them to 5 clusters.

**Figure 2. Inserting a signature size value into each MIMOSA key.** In this example, different parameters are specified. Signatures may have 3 or 4 elements; and two signatures are judged similar if their Jaccard similarity is $\geq 0.4$. The MIMOSA instance receives signatures A-B-C-D, E-F-G, A-B-E-F. **(A)** At $t = 3$, two matching keys (A-B and E-F) are found, so clusters #1 and #2 are candidate clusters for signature A-B-E-F. However, the Jaccard similarity of A-B-C-D and A-B-C-D is 2/6, which is $< 0.4$. So cluster #1 is a false match. **(B)** A slightly different method of creating keys is used, to eliminate false matches by taking advantage of signature size. The MIMOSA instance inserts into each key an additional value: a signature size. When marking keys, MIMOSA uses the size of the signature received. When checking keys for a match, MIMOSA uses the size of a signature that could be similar to the signature received. Because A-B can be in common between A-B-E-F and a similar signature of size 3 but not size 4, the MIMOSA algorithm checks 3-A-B but not 4-A-B. Thus it does not falsely detect a match to cluster #1.

**Figure 3. Clustering benchmark time measurements, MIMOSA vs. standard centroid algorithm.** The centroid run was stopped after 250,000 data items, because of its excessive runtime, more than a full day. **(A)** Log-log plot of clustering time per data item, as a function of the number of data items received. **(B)** Linear plot of cumulative clustering time, as a function of the number of data items received. The inset magnifies the plot for the first 100 data items.

**Figure 4. Distribution of cluster sizes.** Three log-log histograms of cluster size are shown, for the standard centroid run after 250,000 data items, the MIMOSA run after 250,000 data items, and the MIMOSA run after 10,000,000 data items. Each histogram plots the number of clusters having a given number of members. The results show that many clusters (news topics) have only a small number of members (news articles); and conversely, only a few clusters have a very large number of members.

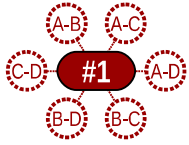
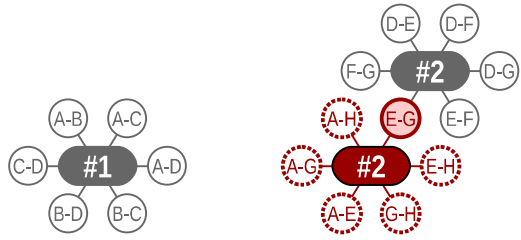
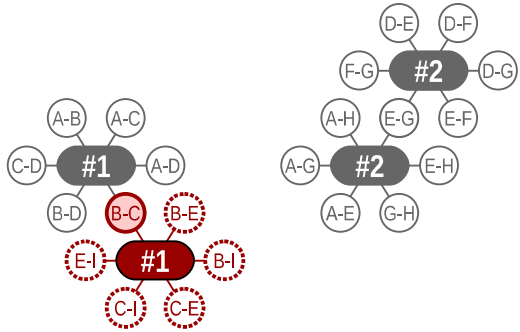
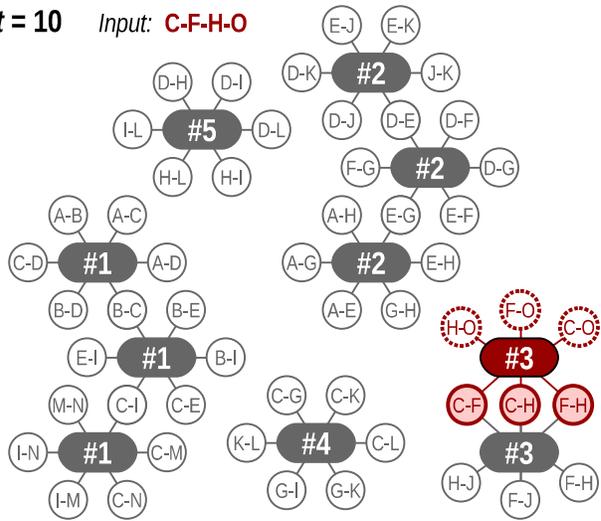

**FIGURE 1**

**FIGURE 2**

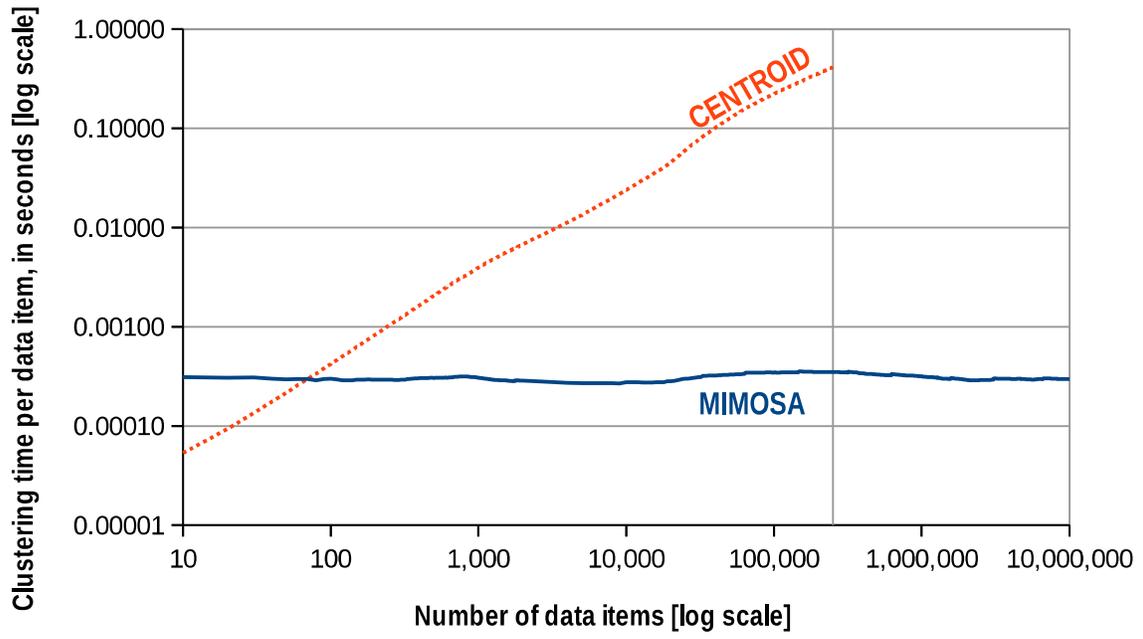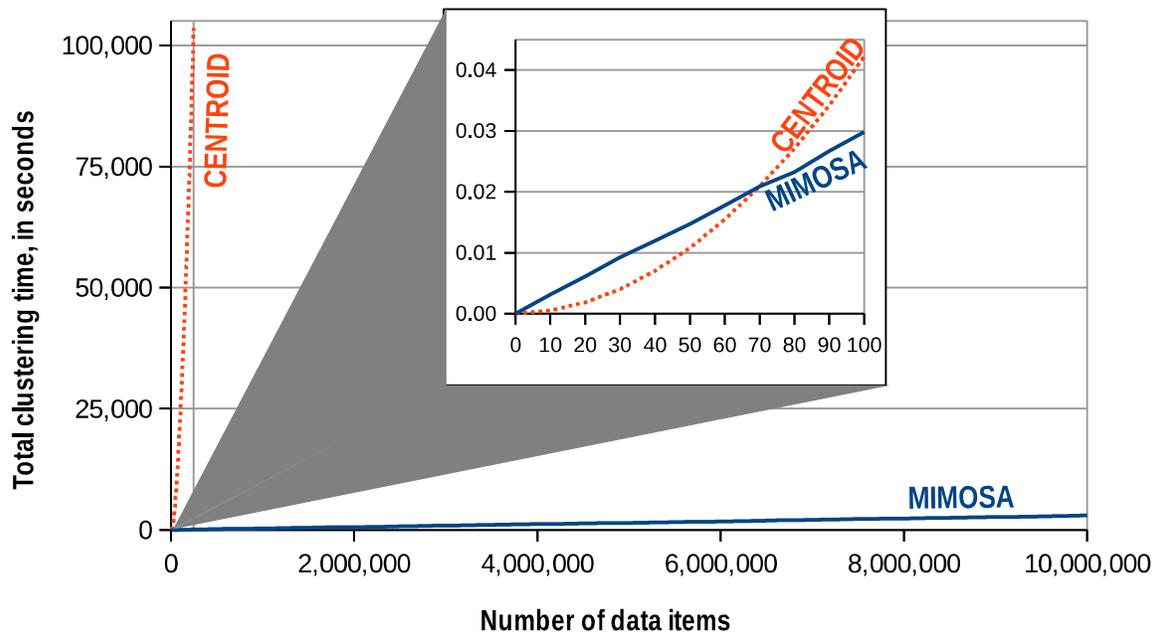

**FIGURE 3**

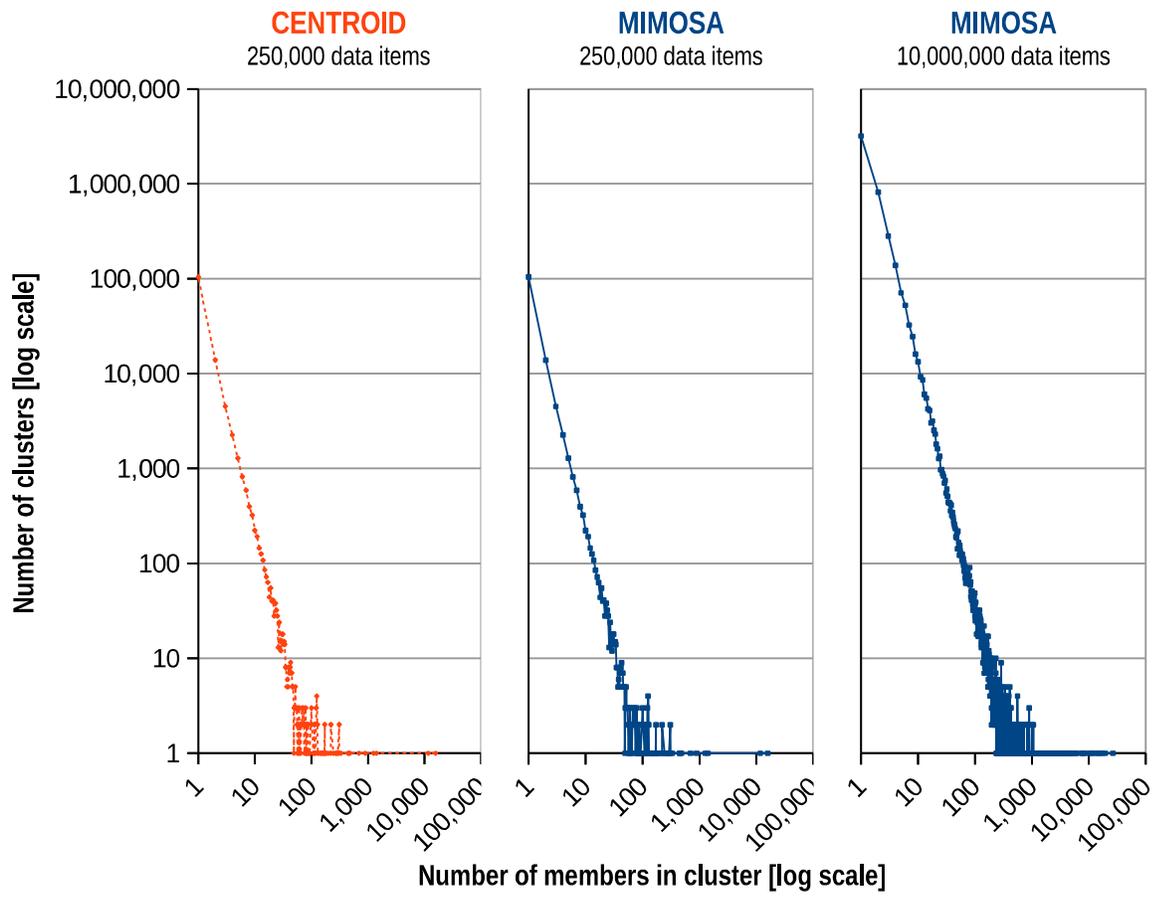

**FIGURE 4**

## SUPPLEMENTARY MATERIAL

Complete data files and software source code, sufficient for full replication, are available publicly in the repository at mimosa.acquiremedia.com, under the dataset1 directory.

Software source code file

    Program: MIMOSA and Centroid implementations        cluster-benchmark.pl

Databases as zipped archives

    Input data: 10,000,000 data item signatures            in-signatures.txt.gz

    Output data: 250,000 Centroid cluster assignments       out-centroid.txt.gz

    Output data: 10,000,000 MIMOSA cluster assignments     out-mimosa.txt.gz

## Materials and Methods

Materials – software

We prepared an implementation of a MIMOSA clustering algorithm and a standard centroid clustering algorithm. The implementations were written in the Perl programming language. The program used a single execution thread. The program source code is available as file S1.

We implemented the following features in both clustering algorithms. (Other variants are possible, but we chose these for simplicity and speed.)

- The first-assigned signature in a cluster is designated the centroid.
- Subsequent signatures matching the centroid are assigned to the cluster, but do not enlarge the cluster neighborhood.
- When an input signature matches more than one centroid, the input is assigned to the earliest (lowest-numbered) corresponding cluster.

The program recorded the results in the output files, archived as S3 and S4, using the following format:

- Column 1: The cluster ID to which the signature was assigned
- Column 2: The ordinal number of the signature
- Column 3: The elements of the signature

In addition, The program was instrumented to record an elapsed time stamp in the output files after every 1,000 signatures (every 10 signatures during the first 2,000). These time stamps are plotted in Figure 3.

Materials – computer platform

We ran the program on a Dell server computer having 96 Gigabytes of random-access memory, and 8 Intel Xeon E5-2470 processing units running at 2.30GHz, under the Centos 6.4 operating system using the Linux 2.6.32 kernel.

Materials – input data

The input data were signatures derived from 10,000,000 news articles collected by NewsEdge.com. Each article, comprising text and metadata, was processed by Acquire Media's proprietary Metabot semantic analyzer program, which identifies entities such as company names, stock trading symbols, person names, and location names, as well as subject and industry taxonomic codes inferred from the content. Metabot's output includes a signature for each story, comprising up to 12 keyword elements, in a lexicographically sorted order.

We truncated signatures that had more than 10 elements, by removing the last ones in excess of 10. The resulting signatures each had between 2 and 10 elements.

We anonymized the elements of each signature by substituting random strings of equal length, while preserving matches, and re-sorting the elements. We put each signature on a line, using hyphens to separate elements. The full set of input signatures is available as zipped archive S2.

Methods

We ran the MIMOSA and centroid benchmarks consecutively, not concurrently. No other applications were run during the benchmarks. The full commands to run the program were:

```
./cluster-benchmark.pl -t 0.6 -s 2-10 -c 250000 <in-signatures.txt >out-centroid.txt
./cluster-benchmark.pl -t 0.6 -s 2-10 -m 10000000 <in-signatures.txt >out-mimosa.txt
```